# Charge separation by standing waves of strongly focused single-cycle terahertz pulses for particle acceleration

Szabolcs Turnár[1,3*], Zoltán Tibai[1], László Pálfalvi[1,3], Csaba Korpa[1], Gábor Almási[1,2], and János Hebling[1,2*]

[1]Institute of Physics, University of Pécs, 7624 Pécs, Hungary
[2]Szentágothai Research Centre, University of Pécs, 7624 Pécs, Hungary
[3]HUN-REN–PTE High-Field Terahertz Research Group, 7624 Pécs, Hungary

*turnarszabolcs@fizika.ttk.pte.hu
*hebling@fizika.ttk.pte.hu

**Abstract:** Electron and ion acceleration in laser beam driven gas and solid-state plasmas are well-known and thoroughly investigated. Here we propose and numerically investigate an ion acceleration scheme using the Coulomb explosion of ion bunches created by separating electron and ion bunches from gas plasma by the standing wave of counter-propagating THz/laser pulse pair(s). The low density and the supposed only few-cycle long THz/laser pulses ensure that much less complicated mechanisms play a role in the separation and acceleration processes, resulting in more predictable outcomes than for single-laser-beam-driven solid-state plasmas. The charge separation is investigated using the EPOCH PIC code. After the charge separation, the ion dynamics ar followed by the GPT code. Calculations predict the generation of relativistic electron bunches and ion bunches with sub-MeV average energy, and with distribution suitable for very efficient use of the accelerated ions in ultrashort-pulse neutron sources.

**Introduction**

Laser-driven electron and ion acceleration is a hot topic in laser-plasma physics [1-4]. Although acceleration to higher than 1 GeV electron energy has been demonstrated in plasmas driven by (sub-) PW lasers [1-4], there is a strong interest in acceleration schemes producing only sub-relativistic or weakly relativistic electrons, which would need driving lasers having orders of magnitude smaller peak-power. For such purposes, dielectric grating or dielectric lined waveguide structures have been suggested and investigated [5-7]. Using THz pulses instead of near-IR laser pulses is more promising because the longer wavelength and oscillation period make the acceleration of electron bunches with larger size and charge possible and relax the needed timing accuracy between electron source and accelerator driving field [8].

In the last few decades, several laser-based ion acceleration mechanisms have been suggested and extensively studied: the target normal sheath acceleration (TNSA), the collision-less electrostatic shock acceleration (CESA), the radiation pressure acceleration (RPA), and the Coulomb explosion acceleration (CEA) [9]. Although a somewhat successful CESA acceleration of ions in gas plasma has been demonstrated [10], thin metal or plastic foil targets are used in most laser-ion acceleration experiments. In this respect, the CEA mechanism is another exception besides the CESA. This method was most extensively investigated in cluster plasmas [11,12]. If other mechanisms are insignificant, CEA in clusters results in omnidirectional acceleration. Although such property is disadvantageous for many applications, it does not mean a significant disadvantage in such important applications as, for example, ultrashort pulse neutron generation by DD fusion in deuterium (D) clusters [13]. In the laser ion acceleration schemes the laser primarily accelerates the electrons, and the ions are accelerated in a large part by the created strong electric field between the electrons and ions.

Here, we propose and numerically investigate a not yet explored laser-driven charge separation scheme, the standing-wave driven charge separation (SWCS, see Fig. 1): using counter-propagating driving THz/laser pulse pair(s) to separate the electrons and ions in the gas plasma. Because of the symmetry of the arrangement, the resultant momentum of the driving pulses is zero, and the opposite sign momenta of the electron- and ion bunches are created entirely by the electric field of the driving pulses. Because of the small mass of the electrons, they gain significant, even relativistic energy during the charge separation, while the ions remain almost standing. The ions gain energy later by transforming the electrostatic potential energy of the separated ion bunch to kinetic energy by the Coulomb explosion (CE). The low density of the plasma and the supposed only few-cycle long laser pulses ensure that much less complicated mechanisms play a role in the acceleration process, resulting in a more predictable outcome than the (single-beam) laser-driven solid plasmas. Furthermore, the low density leads to the proposed scheme's more stable and controllable functioning. The proposed SWCS setup and the alternative solution for strong focusing by using paraboloid ring are shown in Fig. 1a and Fig. 1b, respectively.

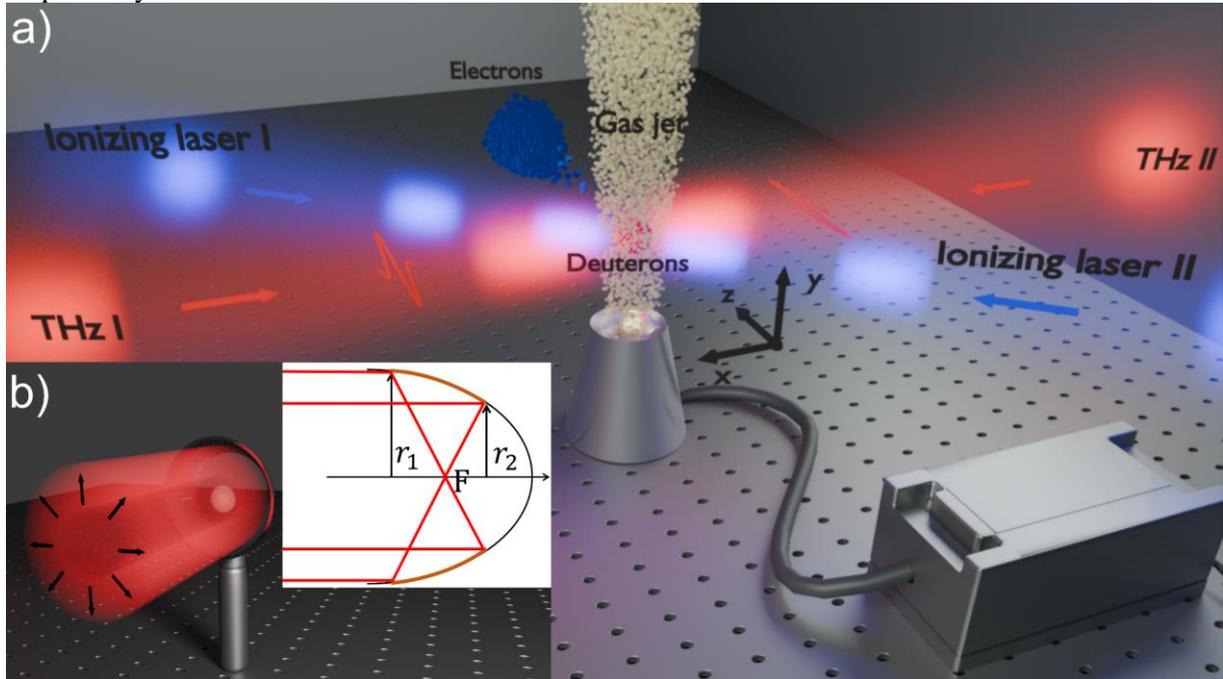

Figure 1. The proposed versions of SWCS setup. a) The standing wave is created by (1-3) pair(s) of counter-propagating near single-cycle focused THz/laser pulses. b) The standing wave is created by focusing a beam by a paraboloid ring.

For both cases, it is supposed that the plasma is generated inside a vacuum chamber by multiphoton ionization of an appropriate gas using a pair of laser beams, delivering ultrashort pulses with low energy but high intensity. Directly after the ionization, the average energies of electrons and ions are only of the order of eV, many orders of magnitude smaller than their final energy. In one of the proposed setups, illustrated in Fig. 1a, 1 – 3 pairs of counter-propagating focused THz beams create the standing wave used to separate the electrons and the ions. In the other proposed setup, illustrated in Fig. 1b, a ring segment of a paraboloid mirror is used to focus one THz/laser beam having radial polarization. Details of the experimental setups are described in the subsection Methods: Possible experimental setups.

The first stage of the processes, the SWCS, is simulated by the EPOCH PIC code, which traces the movement of the electrons and ions and the development of the electromagnetic field created by the separation and acceleration of the charges. In the second stage, the CE of the ion bunch is simulated by the GPT code up to 1-5 ns after separating the electrons and ions. Details of the EPOCH and GPT codes and their applications are described in the subsection Methods: Application of the EPOCH and the GPT codes. According to the simulations, generating deuterons with sub-MeV energy and 12.5 nC bunch charge is possible using two pairs of 0.3 THz pulses with 25 MV/cm electric field amplitude each.

**Results**

Most model calculations were performed for the arrangement shown in Fig. 1a, using deuterium gas and assuming the following parameters. The gas pressure was 2.07 Pa, and the laser beams traveling opposite to each other ionized the gas in a cylinder with a diameter $D$=660 μm and a length $h$=224 μm, in which the total charge of electrons and ions was 12.5 nC. Each of the four THz beams was focused to a field/intensity *FWHM* diameter of 1610 µm/1140 µm (1.61x$\lambda_{THz}$/1.14x$\lambda_{THz}$), the THz pulse frequency was 0.3 THz, and the wavelength $\lambda_{THz}$ = 1000 µm. Each of the four THz pulses was "single-cycle" in terms of field strength (the FWHM duration equals the 3.33 ps period time), 0.71 cycle in terms of intensity, and the peak electric field strength at the focus (center of the plasma) was 25 MV/cm.

**Standing-wave driven charge separation (SWCS)**
Figure 2a shows the energy-colored spatial distribution of electrons and ions for the above parameters at 4.2 ps after the plasma–THz interaction. Due to their large mass (compared to electrons), the deuteron ions still occupy (essentially) the original volume. Even on the right side (where the positive excess charge first develops due to the electrons moving to the left), the ions still have only a few keV of energy, which they acquire through Coulomb repulsion. In contrast, the electrons have already moved about 1 mm. Some have been accelerated to relativistic speeds by the THz field (and, to a lesser extent, by the Coulomb repulsion) and have energies of the order of MeV.

Since the two most essential characteristics of the separated (and accelerated) electron and ion bunches, from the point of view of applications, are the number of accelerated particles (or the proportional charge of the electron or ion bunch) and the average energy of the electrons or ions, we examined how the maximum charge of the electron and ion bunch and the average energy of the electrons and ions depend on the field strength of the applied THz pulses for the parameters given in the previous paragraph. The result is shown in Fig. 2b. In the 0 - (4x)25 MV/cm THz field strength range, the charge of the separable electron or ion bunch depends superlinearly on the THz field. At higher field strengths, the charge is nearly proportional to the field. The average ion energy shows the same THz field strength dependence as the charge. Fig. 2c. shows the electron and ion acceleration efficiency, i.e. the ratio of the kinetic energy of the electron or ion bunch to the total energy of the driving THz pulses as a function of the THz field. These curves show a rapid increase in the 0 - (4x)25 MV/cm range and become nearly constant above 4x25 MV/cm.

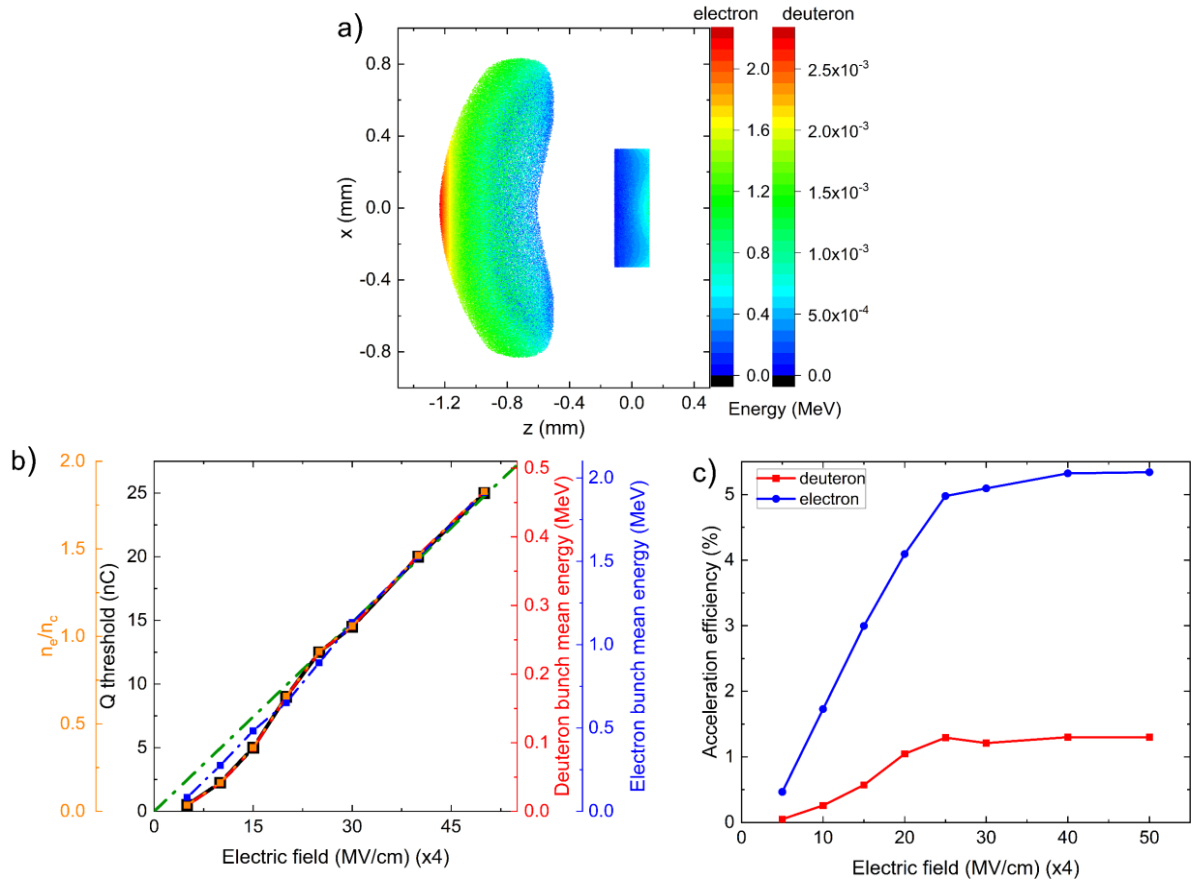

Figure 2. Separation of the electron- and ion bunches. a) Spatial distribution of electrons (at $z < -400$ µm) and ions (around $z = 0$ µm) at 4.2 ps after starting the plasma – THz field interaction. Please notice the near three orders of magnitude difference between the energy scale of the electrons and ions (The 3D plot of the electron distribution is shown in Supplementary Fig 1). b) Dependence of the maximal separable charge and the average electron- and ion energy at 4.2 ps and 5 ns, respectively, after the plasma – THz field interaction (the electron density normalized to its critical value is also indicated). c) The acceleration efficiencies (see text) versus the peak electric field of the "single-cycle" terahertz pulses. (The total electric field is four times larger.) The application of the setup sketched in Figure 1a, but with four beams (two pairs of counter-propagating beams), with the indicated field strengths was supposed.

The results presented above were obtained assuming that the THz pulses generating the standing wave are "single-cycle", i.e., they contain only 1.00/0.71 time periods within the full width at half-maximum of the field strength/intensity envelope. Such THz pulses can be produced by optical rectification of ultrashort laser pulses at the assumed 0.3 THz frequency. However, THz pulses with higher average frequencies (> 1.0 and > 3 THz, respectively, when using lithium niobate or GaAs) contain more oscillations. Furthermore, the possibility arises (see Discussion and Concluding remarks) to use a standing wave generated directly by laser pulses for charge separation. However, the shortest laser pulses typically contain ≈ 2 optical periods within the intensity half-maximum width. Therefore, we investigated whether multi-cycle pulses can be used for charge separation. Figure 3 shows the spatial distribution of electrons at $1^{1}/_{4}$ period (4.2 ps) after the plasma-THz interaction for "single"-, two-, and three-cycle THz pulses. The electron-ion separation is perfect for a "single-cycle" THz pulse. During the examined time, all electrons move away from the ions, located around the $z = 0$ mm position, by at least half a mm. When multi-cycle THz pulses are used, this distance and the average energy of the electron bunches decreases with increasing pulse duration, and a non-negligible part of the electrons only temporarily moves away from the ion bunch, remaining in its vicinity for a longer period. Although the ions can also gain significant energy, however a substantial part of them recombines with the electrons before reaching the far zone.

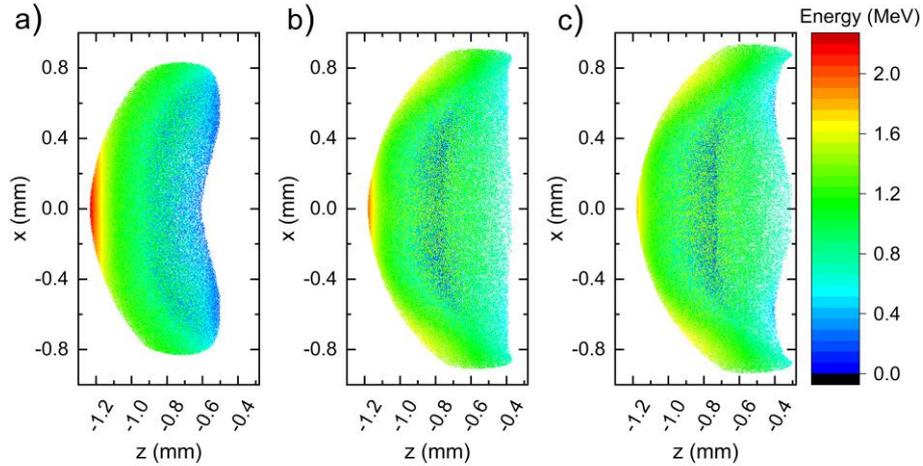

Figure 3. The importance of a "single-cycle" character for the driving THz field. Spatial distribution of electrons at 4.2 ps after the interaction of the plasma and the THz field having a) "single"-, b) two-, and c) three-cycle temporal shape.

**Electron acceleration**

As shown in Figures 2 and 3, the THz standing wave used for spatial separation of electrons and ions accelerates the electrons to sub-relativistic or even relativistic velocities at the investigated field strengths, i.e., the investigated arrangement can be used as an electron gun to produce (sub-) relativistic ultrashort electron bunches. The maximum electron energy gets only a slight contribution from the Coulomb repulsion in the electron bunch. This phenomenon is demonstrated in Figure 4, which shows the spatial distribution (4a and 4b) and energy spectrum (4c and 4d) of the electrons for plasma charges of 1 pC (4a and 4c) and 10 nC (4b and 4d). (The other parameters are set to their default values.). The average energy of the accelerated electron bunch and its spatial distribution depend only slightly on the charge. By varying the charge of the bunch over three orders of magnitude, the brightness of the bunch is approximately proportional to the charge (see the inset of Fig. 4d). Only at the highest charge of 10 nC does strong Coulomb repulsion occur, drastically increasing the emittance of the bunch and reducing its brightness.

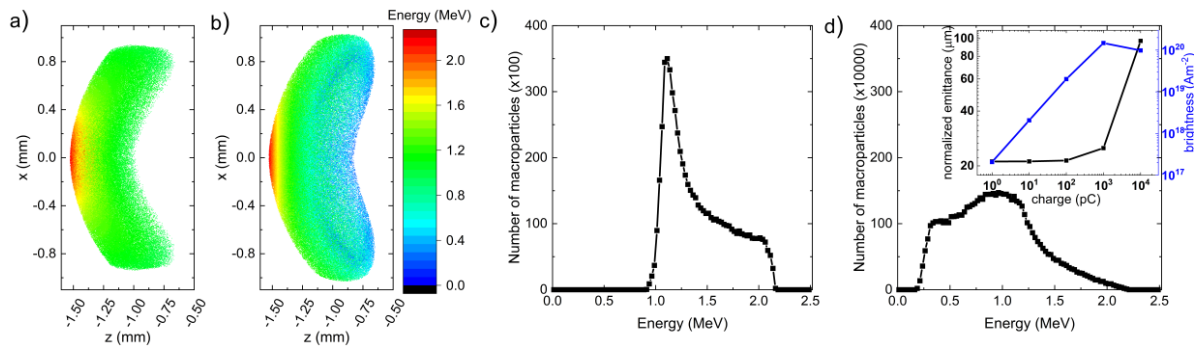

Figure 4. Acceleration of electron bunches by SWCS from laser-generated plasma. Spatial distribution of accelerated electron bunches at 4.2 ps after the interaction of 4 "single"-cycle THz beams having 25 MV/cm maximal field each with the plasma for a bunch charge of a) 1 pC and b) 10 nC. c) - d) Energy distribution of the electron bunches for the same cases. The charge dependence of the bunch emittance and brightness are shown in the inset of figure d. One macroparticle corresponds to approximately 10 and 1000 individual electrons in the case of 1 pC and 10 nC total charge, respectively.

This article does not intend to optimize the electron gun based on standing waves. Still, we note that the emittance can be reduced by more than two orders of magnitude by 5 (125) times reducing the size (volume) of the initial plasma. Both the emittance and the energy spread depend on the initial shape of the plasma (see Supplementary Fig. 2 and Fig. 3).

**Ion acceleration**

Figure 5 shows the energy-colored spatial distribution and energy spectrum of the deuteron bunch created from deuterium gas or heavy water vapor plasma by charge separation using THz SWCS, 1 ns after the plasma-THz pulse interaction. Since the deuteron is 3670 times heavier than the electron, the center of gravity of the deuteron bunch acquires only a small velocity, and the kinetic energy of the deuterons (unlike the case of electrons) is mainly caused by the Coulomb explosion of the deuteron bunch. Therefore, as shown in Figures 5a and 5b, 1 ns after the separation of the electrons, the deuteron bunch has an approximately spherical shape, and the spatial distribution of the energy is also nearly spherically symmetric. The slight deviation from spherical symmetry (different maximum energies in the $x$ and $z$ directions) is determined by the initial plasma shape. The small asymmetry in $+/-z$ direction, considering the energy and the 0.52/0.48 ratio of the ions moving in the $+/- z$ direction, is caused by the direct acceleration effect of the THz standing wave.

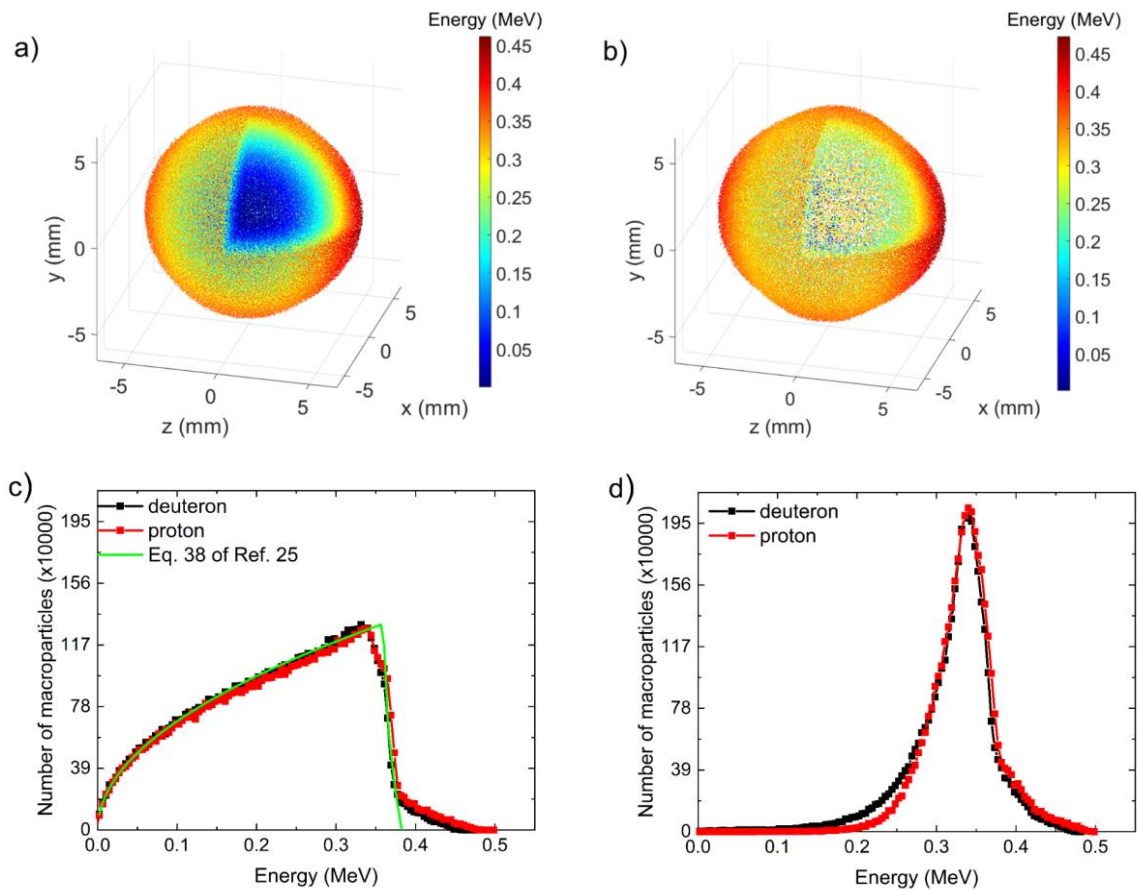

Figure 5. Spatial distribution (a-b) and energy spectra (c-d) of the accelerated deuterons at 1 ns after the interaction of the THz pulses with a), c) deuterium, and b), d) heavy water plasma. In Figures 5a and 5b, the ion energy is represented by a color scale. 4 THz beams with 25 MV/cm maximal field each and a bunch charge of 12.5 nC are supposed. (In Figures 5c and 5d, in addition to the deuteron spectrum (black), the proton (red) spectrum is also shown for hydrogen (Fig. 5c) and water (Fig. 5d) plasmas.

According to the energy spectra presented in Figures 5c and 5d, the maximum deuteron/proton energy is essentially the same in the two examined cases, with a value of approximately 0.47 MeV. However, the average energy of the deuterons/protons is significantly higher in the case of heavy water/water plasma (0.32 MeV) than in the case of deuterium/hydrogen gas plasma (0.22 MeV), and the energy distribution is narrower. In the latter case, however, the charge of the ion bunch is twice as large, which is advantageous for many applications.

**Neutron generation**
The ions with the predicted more than 100 keV energies could be suitable for neutron generation either by the d + d →n + $^3$He or the t + d →n + $^4$He nuclear reactions. Although the t + d →n + $^3$He reaction has a larger cross-section, due to the easier availability of deuterium, as an example we consider neutron generation by accelerated deuterons interacting with deuterium medium where they essentially come to rest. For the cross-section of the relevant reaction, d + d →n + $^3$He, simple parameterizations are available [15,16], and they agree to better than 4 % for projectile-deuteron-kinetic-energy *E* less than 1 MeV, which is the region of interest in our simulation. In the following, we use the parameterization given in Ref. [15], which produces smaller cross-section values for E>0.23 MeV and slightly larger ones below that value. The other quantity needed for neutron yield calculation is deuteron's stopping power S(E) in deuterium medium. In the apparent absence of measured results for that stopping power, one can resort to a phenomenologically improved theoretical formula [17] for arbitrary ion and medium, which reasonably fits several measured results. Alternatively, one may start from a phenomenological fit to stopping the proton in hydrogen [18], and based on that, obtain an approximate stopping power of deuteron in deuterium. One can achieve that transition using the fact that the stopping power of non-relativistic ions of different mass, but the same charge and velocity is equal to high accuracy, the relative difference being of the order of $m_e/m_l$, with $m_e$ and $m_l$ denoting the mass of electron and lighter ion, respectively [17]. Also, the difference of stopping by electrons in hydrogen, $^1$H, and deuterium, $^2$H, is almost negligible since the ionization energies are very close. Comparing the two approaches, one sees that the obtained stopping-power values are better than 3% for deuteron kinetic energy between 0.2 MeV and 1 MeV. For energies less than 0.2 MeV, the formula of Ref. [17] gives significantly larger values by up to 60%. We consider the careful low-energy fit to a proton in hydrogen of Ref. [18] much more reliable than the theoretical formula, which is known to be problematic at low energies, and take the deuteron in deuterium value based on the fit for our yield computation.

The yield (probability of reaction) per incoming projectile of energy *E* is given [19] as

$$y(E) = \int_0^E \frac{\sigma(E')}{S(E')} dE' \qquad (1)$$

where σ is the reaction cross-section, and the stopping power is

$$S(E) = -\frac{1}{n}\frac{dE}{dx} \qquad (2)$$

with *n* being the target atoms' density (number per unit volume). The y(*E*) yield function is calculated according to Eq. 1, and the procedure presented above is plotted in Fig. 6.

The ⟨y⟩ average yield for deuterons with energy distribution function *P(E)* extending until energy $E_{max}$ and normalized to unity is simply

$$\langle y \rangle = \int_0^{E_{max}} P(E) y(E) dE. \qquad (3)$$

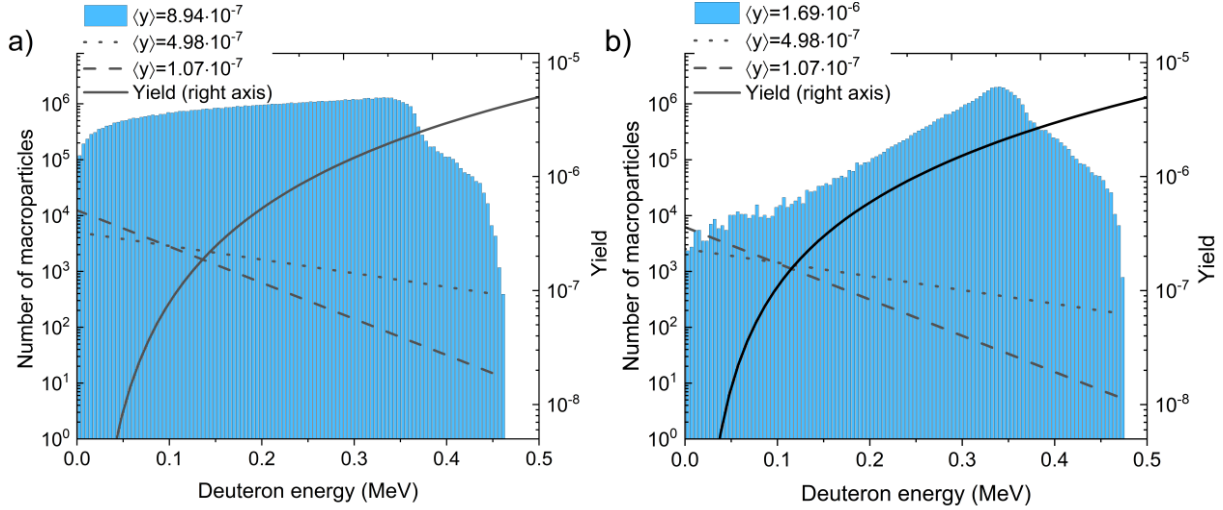

Figure 6. Calculated neutron yield. The calculated neutron yield versus deuteron energy of the d + d →n + $^3$He reaction (solid line), the energy spectra of the deuterons obtained with the CE of a) deuterium and b) heavy water plasma (the identical distributions as shown in Fig. 5c and d, plotted here on a logarithmic scale, bar charts), and the energy spectra having the slope as of the experimental spectra reported in Refs. 20 and 21 (dashed and dotted lines).

We show the average yields for deuterium in Fig.6a and for heavy water plasma in Fig 6b. The bar charts show the deuterium ion energy distributions (through the relevant macroparticle distributions). At the same time, the straight lines correspond to exponentially decreasing functions with the cut-off energies equal to those of bar charts. The dashed lines use the slope of the experimental distribution shown in Ref. [20], and the dotted lines have the slope of fitted exponential in Ref. [21].

For the energy distributions obtained by our simulation of the CE process, the $\langle y \rangle$ average yields turn out to be 8.94·10$^{-7}$ and 1.69·10$^{-6}$, respectively, while for the two exponentially decreasing distributions with equal cut-off energies of 0.47 MeV, it becomes 4.98·10$^{-7}$, and 1.07·10$^{-7}$, respectively. Multiplication of the above given average deuteron yields for the energy distribution shown in Fig. 5a and 5b, with the number of deuterons (7.8 x 10$^{10}$) corresponding to the 12.5 nC ion bunch charge predicts generation of 6.9 x 10$^4$, and 6.6 x 10$^4$ neutrons, respectively. We notice that, because of the larger reaction cross section, generation of nearly two orders of magnitude larger number of neutrons predicted with the same deuteron bunch if the t + d →n + $^4$He nuclear reaction is used.

## Discussion

Before discussing the results, we note that we verified the EPOCH code by tracking the momentum during and after the charge separation. Figure 7 displays the *z* components of the momentum of the electron and ion bunches, and the momentums of the electromagnetic fields around the electron and ion bunches, respectively. In these calculations, a five times smaller initial plasma with a five times smaller charge was supposed as compared to the default. Four pairwise counter-propagating THz beams with 25 MV/cm peak field were supposed. All four beams have a 1.87·10$^{-10}$ Ns momentum pointing into the *x*, -*x*, *y*, and –*y* directions, respectively, but together they have null momentum.

Similarly, the momentum of the electrons and ions is also null before the interaction with the THz beams (which starts at 6.7 ps). During the interaction (and the charge separation), electrons and ions gain about ±2·10$^{-11}$ Ns momentum. These momentums (with good accuracy) are parallel with the oscillating electric field of the THz beams. The *x* and *y* components of the momentums (not shown) are two orders of magnitude smaller than the *z* components.

After the charge separation (until about 10 ps in Fig. 7), a fast decrease in both the electron and ion bunch momentum appears. The reason for this is the pulling Coulomb forces between the electron and ion bunches, which are separated but close to each other at that time. This Coulomb force and its effect on the momentums disappear quickly because of the fast increase of the separation of the electron bunch (moving with close to light velocity) and the ion bunch.

The sum of the four investigated momentums should remain zero during and after the interaction of the THz beams with plasma. Although, according to Fig. 7, this is not precisely the case, the sum of the calculated moments remains less than 12 % of the momentum of the deuteron bunch during all of the time investigated. We attribute even this minor discrepancy not to the inaccuracy of the EPOCH code but to the difficulty of the determination of the momentums of the electromagnetic field around the electron and ion bunches (because of the relatively small investigated volume compared to the transversal and longitudinal sizes of the THz beams).

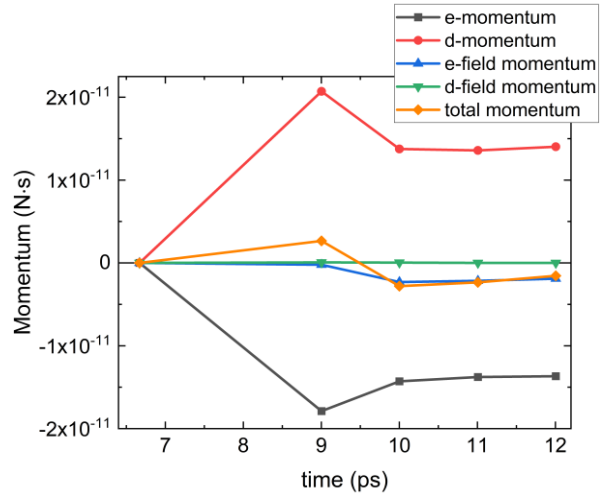

Figure 7. The time dependence of the $z$-components of the moments after the interaction of 4 "single"-cycle THz beams having 25 MV/cm maximal field, each with a plasma with a diameter of 132 μm and a length of 44.8 μm and a bunch charge of 2.5 nC.

Investigation of the charge separation has been performed up to 4 x 50 MV/cm (see Fig. 2). Although producing THz pulses with few tens of MV/cm peak electric field at 0.3 THz frequency is presently challenging, it is expected that using the new generation LN THz sources [14,22] after their full development, it will be possible in the near future. However, there is an upper limit for the THz field as long as the presented idea and calculation method for the charge separation hold. Namely, the THz field will ionize the gas independently of a UV ionizing laser pulse above a given field strength. Although, even in this case, the charge separation is possible, expectedly, it will be less effective, and our model becomes inapplicable. We can estimate this limiting field by applying the tunnel ionization formulas valid for hydrogen (or deuterium) atoms [23]. According to this estimate, about 30 % of the atoms are ionized in half a cycle of a THz pulse having 0.3 THz frequency and 200 MV/cm peak field. Because of the strongly nonlinear character of the tunneling, in the case of 250 MV/cm peak field, close to 100 % of the atoms will be ionized in less than 1/10 temporal period. So, in the case of 0.3 THz "single-cycle" pulses, 200 (4x50) MV/cm field can be considered as the upper limit of the applicability of our simulation method.

According to Figure 2b, in the THz field strength range of 0 – (4x) 25 MV/cm, the charge of the separable electron and ion bunches depends superlinearly on the THz field, but for higher field strengths, the charge is nearly proportional to the field. It is important to notice that the plasma electron density reaches its critical value around the border separating the two dependences (see the leftmost scale in Figure 2b). Since THz radiation cannot penetrate deeply into the plasma above the critical electron density, it has more difficulty in removing electrons from ions. Presumably, this is why the nonlinear field strength dependence changes to a nearly linear dependence.

If we could slowly separate the totality of electrons and ions in the initial plasma as an electron and an ion bunch and move them away from each other, a Coulomb potential energy would build up in both bunches, depending on the size, shape, total charge and charge distribution of the bunches. In the case of a homogeneous sphere, this energy is

$$E_{CPb} = \frac{3}{5} \times \frac{Q^2}{4\pi\epsilon_0 R}, \tag{4}$$

where $Q$ is the charge of the bunch, $\epsilon_0$ is the permittivity of the vacuum, and $R$ is the radius of the bunch. This energy is then converted into the kinetic energy of the electron and ion bunch due to the Coulomb explosion. During the slow separation, work equal to the sum of the potential energies of the two bunches would have to be done, i.e., electrons and ions would be accelerated with 50 % efficiency.

However, such slow separation cannot be achieved with THz standing waves (or in any other way), and according to Fig. 2c, the electron and ion acceleration efficiencies are not greater than ≈5.2 and ≈1.3 %, respectively, even for large fields. There are several reasons for this. The most significant is that the $D \times h$ initial plasma cross-section is much smaller than the cross-section of the THz beams (defined at half the maximum field strength). In the default case, the ratio of these two cross-sections is only 0.073. According to Supplementary Fig. 4a, the sum of the electron and ion acceleration efficiencies strongly depends on this cross-section ratio: a 7-fold (49-fold) change in the linear size (cross-section ratio) of the plasma results in a nearly 5-fold increase in the case of the efficiency. (We did not increase the plasma size further to increase the acceleration efficiency because the number of electrons that could not be permanently detached increases rapidly if the size is larger than the default value.)

Furthermore, the acceleration efficiency also decreased because a small amount (about 1%) of the THz energy turns into the energy of the electromagnetic radiation emitted by the accelerated electron bunch (see Supplementary Fig. 5).

According to Fig. 2c and Supplementary Fig. 4b, the electron acceleration efficiency is significantly higher than the ion acceleration efficiency. This phenomenon occurs because, in the course of the charge separation, the electron and the ion bunch acquire (with opposite signs, but) essentially the same magnitude of momentum (Fig. 7), and due to the much smaller mass of the electrons, this means much larger kinetic energy (even significantly larger than that given by Eq. 4). However, in the case of ions, this excess energy is negligible compared to the potential energy given by Eq. 4.

A formula of the same form as Eq. 4 also gives the Coulomb potential energy in the case of a cylindrical bunch. In our default case, it is (based on numerical calculations):

$$E_{CPdc} = 0.547 \times \frac{Q^2}{4\pi\epsilon_0 R}, \tag{5}$$

(In the case of $D = h$, the coefficient is 0.589 instead of 0.547.) The accuracy of our numerical simulation is supported by the fact that the kinetic energy of the ion bunch calculated 5 ns after the interaction was the same with an accuracy of 1 % as the value calculated based on Eq. 5 (it was smaller).

Dividing Eq. 4 by the number of ions, the average ion energy after the Coulomb explosion can be determined:

$$E_{CPi} = \frac{3}{5} \times \frac{QZe}{4\pi\epsilon_0 R}, \tag{6}$$

where $Ze$ is the charge of the ion. Furthermore, an analytical formula can be derived for the energy spectrum in the case of a uniformly charged sphere [24] or an ellipsoid [25]. In Fig. 5c, the solid green curve shows the energy spectrum calculated using equation 38 of Ref. 25 with the values of $\varepsilon_0 = 0.365$ MeV and $\alpha_0 = 2.94$ applicable assuming an ellipsoid with the charge, volume, and aspect ratio of the default plasma regarding our simulation. This curve closely approximates the spectrum calculated using PIC simulation for a cylinder with the same properties.

The change in the energy spectrum of the light ion upon addition of a heavy (oxygen) ion to the light (deuterium/hydrogen) ion in Fig. 5c and Fig. 5d (increase in the average energy and significant decrease in the energy spread) shows a good qualitative agreement with the expectation based on the analytical formulas presented in Ref. 12. It is important to note that contrary to most laser-plasma based ion acceleration methods, using the SWCS followed by CE the ions mean energy and energy distribution are essentially independent of the ion mass. The slight difference of the proton and deuteron energy spectra shown in Fig. 5c and Fig. 5d is caused by the almost negligible direct acceleration of the ions by the THz field.

Based on the relatively simple processes involved in the SWCS, the formulas describing the average ion energy and energy spectrum following the CE, the equivalence of optical systems having the same linear size/wavelength ratio, and the fact that the work is equal with the force's integral on the path, the following simple scaling rules for unchanged energy spectrum with the THz/laser wavelength are expected: the plasma size ($R$ in the case of a sphere) and the $Q$ charge have to be proportional to the wavelength, and the peak electric field strength of the THz/laser pulse has to be inversely proportional to the wavelength, supposing the same temporal shape (the time scales with the wavelength) of the THz/laser pulse. These scaling rules were verified by comparing our PIC and GPT simulations, changing the supposed THz/laser wavelength by a factor of 1000 (from 1 mm to 1 μm). From these scaling rules, it follows that using the same plasma shape (but not the size), the generation efficiency of accelerated ions is independent of the THz/laser wavelength used. It also follows that the $n_0$ initial electron and ion concentration of the plasma have to be proportional to the inverse square of the wavelength. This means that when using an NIR laser in a SWCS setup, the initial plasma density has to be in the order (or only one order lower than) of the atom density in solids.

**Methods**
**Application of the EPOCH and the GPT codes**
The numerical simulations were performed by two codes: EPOCH [26] (which is a particle-in-cell (PIC) code) and GPT [27] (which is a particle tracer software). The calculation domain of the PIC simulation is 2000 μm, 2000 μm, and 3000 μm in the $x$, $y$, and $z$ directions, respectively, and its centrum is the origo. The x and y axis are parallel with the propagation direction of the two pairs of focused THz beams. The PIC simulations begin utilizing the plasma profile. The plasma electron density was initialized at $n_e/n_c = 0.913$ for $r < 330\,\mu m$ and $|z| < 112\,\mu m$, where $r = \sqrt{x^2 + y^2}$ and $n_c = 1.116 \cdot 10^{21}$ electron $\cdot m^{-3}$. The spatial resolution was 4 μm, 4 μm, and 5 μm in $x$, $y$ and $z$ directions, respectively. Each unit cell had 82 macroparticles, both with negative and positive charge, each representing 1000 electrons or ions, respectively. Both the spatial and temporal profile of the THz pulse was Gaussian by encoding the *profile* and *t_profile* parameters in the laser block of EPOCH with the following parameters: The spatial profile is determined i) at the $x$ boundary as *profile* = $\text{Gauss}\left(\sqrt{y^2 + z^2}, 0, w_0 \cdot \sqrt{1 + \left(\frac{f_c}{z_R}\right)^2}\right)$, and ii) at the y boundary as *profile* = $\text{Gauss}\left(\sqrt{x^2 + z^2}, 0, w_0 \cdot \sqrt{1 + \left(\frac{f_c}{z_R}\right)^2}\right)$, where $f_c$ is the position of the THz beam focus, and $z_R$ is the Rayleigh length. The time profile was determined by using the function of $\text{Gauss}(time, t_c, t_w)$, where $t_c$ is 3.33 ps, and $t_w$ was 1.72 ps. The electric field was given by the $E(r, x, t) = E_0 \frac{w_0}{w(x)} exp\left(-\frac{r^2}{w(x)^2} + i\left\{kz - \omega t + k\frac{r^2}{2R(x)} - \eta(x) + \varphi\right\}\right) \cdot exp(-2ln2(x - ct)^2/c^2\tau^2)$ equation. Gaussian pulses with a frequency 0.3 THz, duration of $\tau = 2.025$ ps, beam waist of $w_0 = 1000\mu m$, peak electric field of 33.33 MV/cm were launched from the left and right boundaries in the $x$ and $y$ directions. The polarization of the THz pulses was oriented in the $z$-direction. The initial phase was $-30°$. The total simulation time using the PIC code was 12 ps. At that moment, the electron bunch became well separated from the ion bunch, and the electric force between the ripped electrons and the remaining ions became close to negligible. The position and velocity values of the deuteron macroparticles at this time were imported into the GPT code, where the process of the Coulomb explosion was investigated until 1-5 ns. During the GPT simulation, we used the SpaceCharge3D function, which takes into account the 3D space-charge effects by using a point-to-point method. The Plummer radius was set to 5 nm.

**Possible experimental setups**

Although this article deals with the operating principle and the numerical simulation of a new arrangement, it is worth reviewing the methods and tools required for its future implementation. In the arrangement shown in Figure 1, a gas jet flowing out of a nozzle is assumed as the initial neutral material. This solution is suitable and practical for the case of charge-separating radiation with a frequency of 0.3 THz, i.e., a wavelength of 1 mm, assumed in most of our numerical simulations, since only 2 Pa of pressure needs to be provided. As shown in the chapter of Discussion, according to one of the scaling laws of the Coulomb explosion, this pressure is inversely proportional to the square of the wavelength. Based on this, the solution presented in Figure 1 can be used in the entire 0.3 – 10 THz frequency range of efficient terahertz sources since, by using appropriate vacuum pumps, it is possible to ensure that the ions and electrons reach the place of their application even at the gas jet pressure of 2000 Pa, that is about 0.02 atm, required for 10 THz.

If we want to use lasers or OPA operating in the near or mid-infrared range directly for charge separation, then in the case of a gas medium, a pressure so high (1 – 20 atm) would be required, at which the generated ions or electrons cannot be utilized. In such a case, a very thin (0.1 - 1 μm) liquid jet could be a solution [28].

**Multiphoton ionization of the gas**

As we have shown in the Discussion, for charge separation using short pulsed standing waves to be truly effective, it is necessary to ionize the initial gas (or liquid or a foil) with an ultrashort laser pulse with a duration significantly shorter than the period of the standing wave used for charge separation at a suitable time synchronized to the standing wave. If THz pulses are used for charge separation, pulses for photoionization can be split out from the pump laser pulses used to generate the THz pulses using a beam splitter. Compressing these pulses in time and generating their third or fourth harmonic is advisable. For example, using 0.3 THz standing waves for charge separation, Yb pump laser pulses with a duration of 0.3 – 1.0 ps can be used for THz pulse generation. The fourth harmonic of these (≈257 nm wavelength) produces a photon energy of about 4 eV. Thus, hydrogen or deuterium with an ionization energy of 13.6 eV can be ionized by the fourth harmonic beam using four-photon absorption. Since four-photon absorption is strongly intensity-dependent (proportional to the fourth power of the intensity), the diameter of the focused harmonic beam determines the cross-section of the plasma generated from the (e.g., hydrogen) gas. The plasma length can be controlled with the duration of the fourth harmonic if two opposing beams are used for ionization. The plasma length of 224 μm, assumed in most calculations, can be generated by opposing ionizing pulses of about 0.5 ps. According to Ref. 29, for a wavelength of ≈250 nm, an intensity of ≈$1.3 \cdot 10^{14}$ W/cm$^2$ is required for efficient plasma generation.

**Generation of the standing wave**

One arrangement for generating the high-field (THz) standing wave for charge separation is shown in Figure 1a. Here, most simply, the standing wave is generated by the superposition of two focused beams (usually obtained by splitting the beam of a THz source) traveling opposite each other. It can be easily shown that for the same total THz energy, a higher superposed field strength can be generated if more pairs of beams are used, intercepted at an angle of 180°/$N_p$ for each other around the z-axis, where $N_p$ is the number of beam pairs. If a beam is split into $N_p$ pairs of equal-intensity beams and superimposed, the maximum field strength produced at the intersection of the superimposed beams is given by the equation

$$E_{Np} = \sqrt{2 \cdot N_p} \times E_{1B}, \tag{7}$$

where $E_{1B}$ is the maximum field strength that could be achieved by focusing a single beam. Based on this, by using 1, 2, 3, etc. beam pairs, a field strength of $\sqrt{2}$, $\sqrt{4}$, $\sqrt{6}$, etc. larger can be achieved if the focusing conditions are the same. However, the large numerical aperture required for strong focusing is impossible for larger than $N_p=3$. In most of the calculations, we assumed using $N_p= 2$ beam pairs. Using multiple beam pairs also has the advantage that the superimposed space becomes increasingly cylindrically symmetric as $N_p$ increases.

A perfectly cylindrically symmetric field distribution for charge separation is provided by the simpler arrangement using a paraboloid ring shown in Figure 1b [30].

**Concluding remarks**

In conclusion, we proposed standing wave charge separation (SWCS) for generating electron and ion bunches from gas plasma by using few-cycle electromagnetic pulses. Assuming pulses with 0.3 THz central frequency and 10 – 100 MV/cm field strength, our EPOCH PIC simulations predicted the separation of 0.1 – 10 nC level charges. During the SWCS, the electrons were accelerated to 0.1 – 3 MeV energy level. The CE of the ion bunch generated by the SWCS can accelerate deuterons or other ions to the sub-MeV level. The energy spectrum of the accelerated deuterons has close to square root dependence, which is more suitable for neutron generation than the exponentially decreasing spectrum usual in laser-ion acceleration. An even better spectrum, having a narrow (+/-10 %) energy distribution, can be achieved using two-component gas plasma. Furthermore, according to our preliminary results, using a two-layer plasma with spatially separated oxygen and deuteron ions, the directed acceleration of deuterons is also possible (see Supplementary Fig. 6). In this case a 0.935/0.065 ratio of the ions moving in the +/- $z$ direction predicted, contrary to the 0.52/0.48 ratio for the homogeneous plasma.

According to simple scaling laws, the shorter wavelength IR or MIR laser pulses can also be used in SWCS setups instead of THz pulses. The important point is that the laser pulses should contain as few optical cycles as possible. Recently, there have been many different efforts to generate such pulses using spectral broadening followed by temporal compression of Ti:sapphire laser pulses [31], coherent combination of carrier-envelop-stable OPA pulses [32], and cascaded intrapulse difference-frequency generation [33], just to mention a few methods. Please note that the needed THz/laser intensity is proportional to the wavelength, but the pulse energy is advantageously inversely proportional to it.

Finally, Supplementary Fig. 7 demonstrates our claim, that in the case of few-cycle pumping, the symmetric arrangement of the SWCS setup ensures a much less complicated charge separation, with a more predictable outcome than using a single laser beam for charge separation of plasmas. Here, the default initial plasma parameters were supposed. Supplementary Fig. 7a and Fig. 7b show the electron distribution after the SWCS process using four beams with 25 MV/cm peak field, and after the charge separation caused by a single beam with 100 MV/cm peak field, respectively. Although in the latter case the used pulse energy is 6 times larger than in the SWCS case, yet the charge separation is not perfect (see Supplementary Fig. 7b), and the movement of the electrons is complicated. Contrary, in the SWCS case, the charge separation is perfect, and the movement of electrons is simple.

# Charge separation by standing waves of strongly focused single-cycle terahertz pulses for particle acceleration

Szabolcs Turnár[1,3*], Zoltán Tibai[1], László Pálfalvi[1,3], Csaba Korpa[1], Gábor Almási[1,2], and János Hebling[1,2*]

[1]Institute of Physics, University of Pécs, 7624 Pécs, Hungary
[2]Szentágothai Research Centre, University of Pécs, 7624 Pécs, Hungary
[3]HUN-REN–PTE High-Field Terahertz Research Group, 7624 Pécs, Hungary

*turnarszabolcs@fizika.ttk.pte.hu
*hebling@fizika.ttk.pte.hu

**Supplementary material**

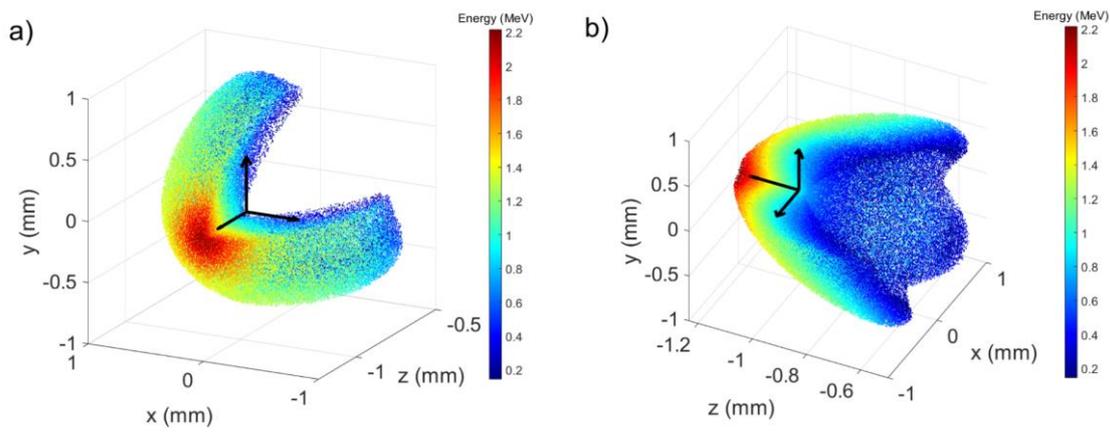

Figure 1. Two perspectives of the same electron distribution as shown in Figure 2a using 3D plots. a) Front view and b) back view. The electron bunch propagates in the –z direction, and a quarter (x<0, y>0) slice is cut out from it for illustration purposes.

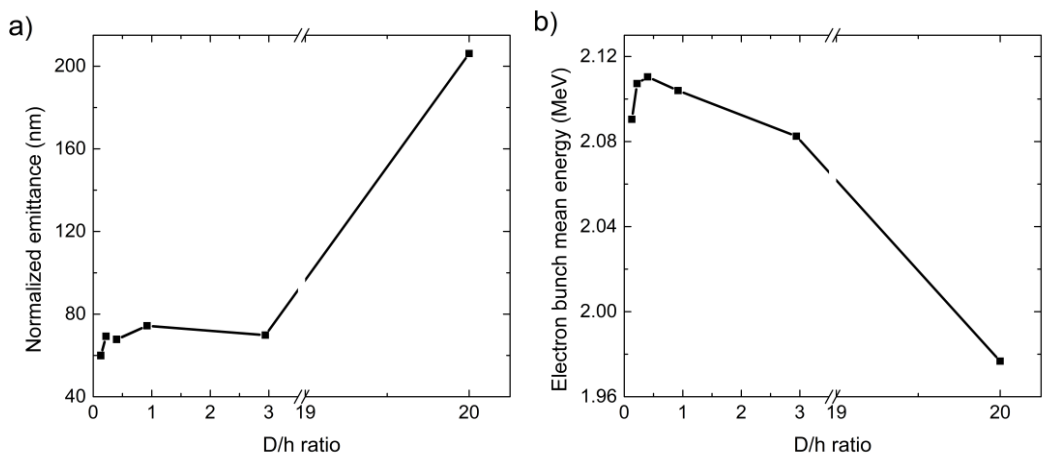

Figure 2. Dependence of the electron bunch properties on the initial plasma shape. Four THz beams with 25 MV/cm maximal field each, a bunch charge of 10 pC and a $6.13 \cdot 10^{-4} mm^3$ initial plasma volume (125 times smaller than the default) was supposed. The charge density was unchanged for all *D/h* ratios. a) The normalized emittance of the electron bunch, and b) the average electron bunch energy versus the diameter-height ratio.

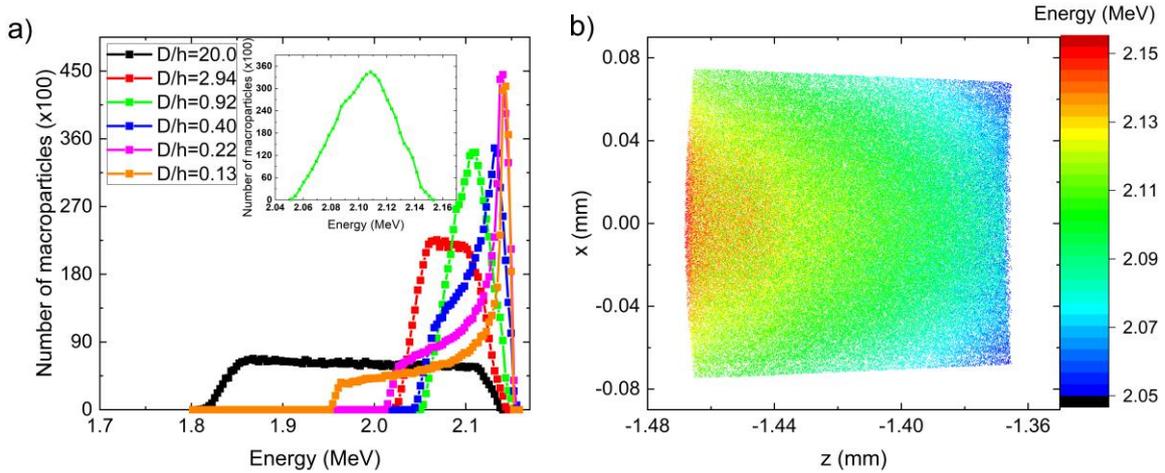

Figure 3. Acceleration of an electron bunch having 10 pC charge and 125 times smaller plasma volume than the default (The conditions are the same as in the case of Figure 2). a) Energy spectra of the accelerated electron bunches for different initial shapes. b) Spatial distribution of accelerated electron bunch at 5.2 ps after the plasma-THz interaction for 0.92 *D/h* ratio, *D*=89.6 μm, *h*=97.2 μm. One macroparticle corresponds to approximately 100 electrons.

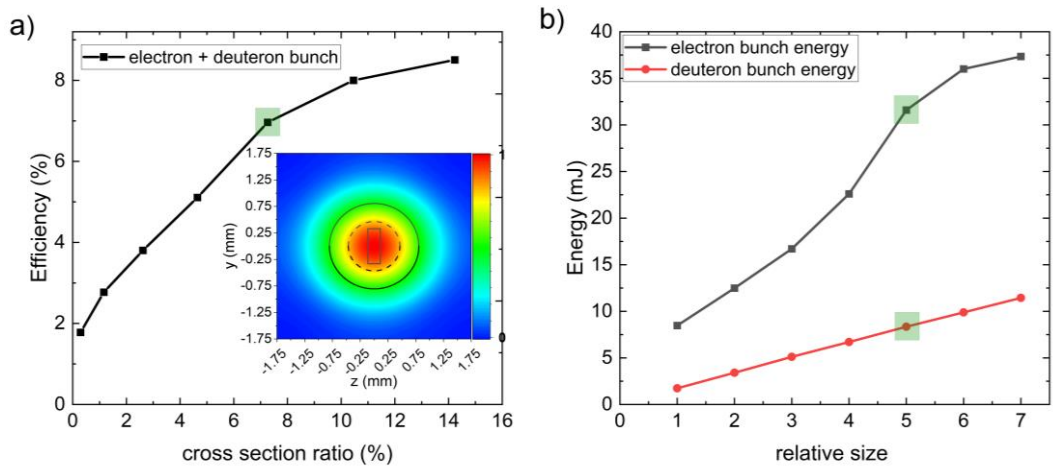

Figure 4. The dependence of the acceleration energy efficiencies of all the particles on the (linear) size of the laser generated initial plasma. The charge was varied linearly with the size. Four THz beams with 40 MV/cm maximal field each were supposed. The smallest investigated plasma size (relative size =1) was 132 μm in diameter and 44.8 μm in length. In this case the charge was 4.3 nC. a) The energy efficiency of the accelerated particles in respect to the cross-section ratio (*D* x *h* initial plasma cross-section divided by the THz beam cross section measured at 50 % electric field of the maximum). Inset indicates the contour plot of the focused electric field. Convention of indications: rectangle = initial plasma in the default case (relative size=5); dashed and solid circles = The electric field is 80 % and 50 % of the maximum electric field, respectively). b) The achieved energy of the electron and deuteron bunches after the acceleration for different initial bunch size. The results for the default size are indicated with green squares in Figure a) and b). The relative size of 5 equals the case of cross section ratio around 7 %.

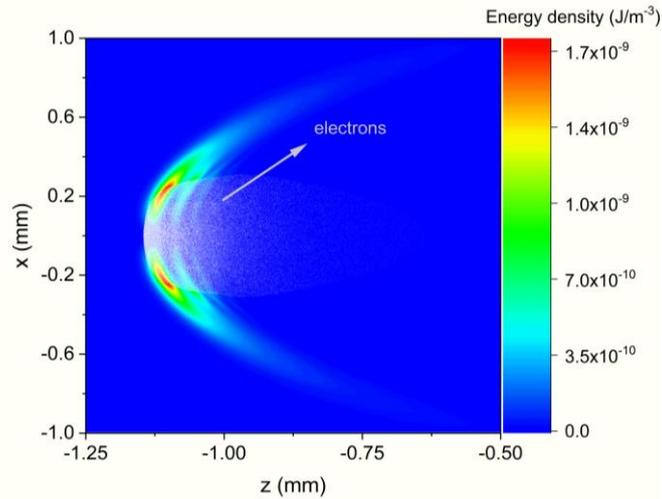

Figure 5. Electromagnetic field energy density around the electrons at 4.2 ps after the plasma-THz interaction. The electrons are marked with gray color. Four THz beams with 25 MV/cm maximal field each were supposed. The investigated initial plasma size was 132 µm in diameter and 44.8 µm in length. The initial charge was 2.5 nC. Integration of the energy density of the electromagnetic field around the electron bunch resulted in 0.76 mJ energy.

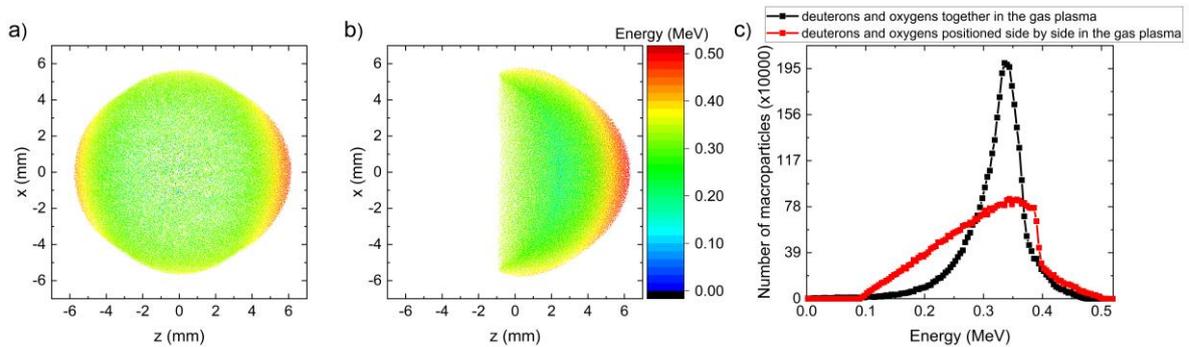

Figure 6. Comparison of the results for two different arrangements. The simulation conditions are the same as the default case except the composition and spatial distribution of the ions. Spatial distribution of the accelerated deuterons at 1 ns after the interaction of the THz pulses with plasma a) when deuterons and oxygens are mixed homogenously and b) when the oxygen and deuterium atoms are in adjacent spaces (the oxygens positioned in negative half part of the initial bunch, while the deuterium are positioned in the positive part of the initial bunch along the z coordinates). c) The energy spectra of case a) and b).

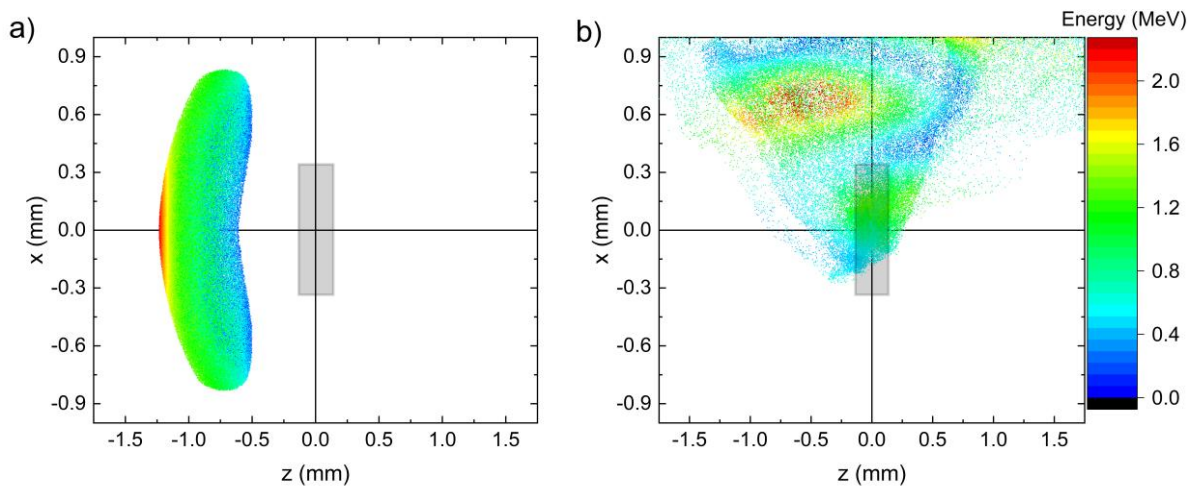

Figure 7. Comparison of charge separation using standing wave and single-beam. The default plasma parameters were used. Spatial distribution of the electron bunch when a) four "single-cycle" THz beams with 25 MV/cm maximal field each generate the standing wave and when b) one piece of three-cycle pulse propagates from the -x direction with 100 MV/cm maximal electric field. The snapshot is created 4.2 and 29 ps after the first interaction, respectively. The gray rectangle represents the deuteron bunch. At the time of the snapshot in b) 70 % of the electrons left the calculation domain.